\documentclass[a4paper,14pt]{revtex4}
\usepackage[english]{babel}
\usepackage{epsfig}
\usepackage{latexsym}
\usepackage{epsfig}
\usepackage{here}
\usepackage{float}
\begin{document}
\title{Density functional theory calculations of the carbon ELNES of small diameter armchair and zigzag nanotubes: core-hole, curvature and momentum transfer orientation effects.}
\author{J.~T.~Titantah$^\dag$, K. Jorissen$^\ddag$ and D. Lamoen$^\dag$\\
{{$^\dag$\small \it TSM}, \small \it  Department of Physics, University of 
Antwerp, Groenenborgerlaan 171, 2020 Antwerpen, Belgium}\\
{{$^\ddag$\small \it EMAT}, \small \it Department of Physics, University of 
Antwerp, Groenenborgerlaan 171, 2020 Antwerpen, Belgium}\\
\date{\normalsize (\today)}}
\begin{abstract}
We perform density functional theory calculations on a series of armchair and zigzag nanotubes of diameters less than 1nm using the all-electron Full-Potential(-Linearised)-Augmented-Plane-Wave (FPLAPW) method. Emphasis is laid on the effects of curvature, the electron beam orientation and the inclusion of the core-hole on the carbon electron energy loss K-edge. The electron energy loss near-edge spectra of all the studied tubes show strong curvature effects compared to that of flat graphene. The curvature induced $\pi-\sigma$ hybridisation is shown to have a more drastic effect on the electronic properties of zigzag tubes than on those of armchair tubes. We show that the core-hole effect must be accounted for in order to correctly reproduce electron energy loss measurements. We also find that, the energy loss near edge spectra of these carbon systems are dominantly dipole selected and that they can be expressed simply as a proportionality with the local momentum projected density of states, thus portraying the weak energy dependence of the transition matrix elements. Compared to graphite, the ELNES of carbon nanotubes show a reduced anisotropy.  
\end{abstract}
\maketitle
\baselineskip=6mm
\section{Introduction}
Since the discovery of carbon nanotubes by Iijima~\cite{iijima91} in 1991, 
much effort has 
been devoted experimentally~\cite{ebbesen92,bethune93,endo93,bougrine99} and 
theoretically~\cite{mintmire92,mintmire95,mintmire98} to study this new material. 
The one dimensional character of single wall nanotubes (SWNT's) enables them to exhibit very interesting physical properties due to the quantum confinement of electrons in a one dimensional lattice. Remarkable optical, thermoelectric, mechanical and electronic properties can be expected. Doping or intercalation ~\cite{baxendale99,pichler99,suzuki98,suzuki01,baierle01,seung02,duclaux02,hsu00,liu03}, structural defects~\cite{meunier01,venema00} and pressure effects~\cite{kazaoui00} are known to enrich these properties. Resonant Raman spectroscopy and optical absorption spectroscopy~\cite{kataura99,kurti99,huong95,pimenta98} have been widely used to describe the electronic structure of isolated and ropes of nanotubes. Scanning tunneling microscopy~\cite{venema00,philip99} has also been used to detect the Van Hove spikes of SWNT's. Most of the theoretical studies are based on the application of the Born-von Karman boundary conditions to the two dimensional graphene 
sheet~\cite{mintmire95,saito98} in a zone-folding technique and the all-valence tight-binding method~\cite{mintmire92,saito98,saito00}. This graphene zone-folding approach is known to be inaccurate to describe the unoccupied density of states (DOS)~\cite{reich02} a few eV's beyond the Fermi level. Density functional theory (DFT) 
calculations~\cite{reich02,suenaga01} have also been  carried out but very few calculations~\cite{suenaga01} probing core-loss spectra are available. Electron energy loss spectrum (EELS) and X-ray absorption spectroscopy (XAS) measurements are available but 
their theoretical interpretation is often limited to the site and angular 
momentum projected local density of states (LDOS). This interpretation does not account for transition matrix elements which are known to be important and sensitive to the momentum transfer direction for anisotropic materials like graphite and carbon nanotubes. While the DOS of a tube is characterised by the Van Hove singularities inherent to the one dimensional nature of the tube, it has been argued that the EEL spectrum is free of any 
such spikes due to the effect of the C1s core-hole in the final state~\cite{knupfer99}. This implies a strong energy dependence of the transition matrix elements. We point out, contrary to this assertion, that the energy dependence of the transition matrix elements is weak meaning that the spikes may always feature in the EELS spectra if the energy resolution is good enough and thermal effects are rendered minimal. The few DFT calculations of energy loss near edge spectra (ELNES) so far available~\cite{suenaga01} do not include the all important core-hole effect on the ELNES. We have included and analysed these effects on the carbon K-edges of nanotubes.   

We perform ab initio calculations on a series of single-wall armchair and zigzag nanotubes.   
We present results on the electronic properties of the carbon 
nanotubes. For comparison with other calculations, we considered the (6,6), (5,0) and (10,0) tubes. The obtained band structure is in good agreement with earlier results for both the (6,6) tube~\cite{reich02} and the (5,0) tube~\cite{li01,machon02}. The calculated C K-edges agree with those of Suenaga et al.\cite{suenaga01} for the (5,0) and (10,0) tubes. This agreement with other calculations validate the method used in this work. Much emphasis is laid on the curvature, orientation and core-hole effects on the C K-edges, as these aspects have not been addressed so far. 

In order to demonstrate the fact that energy loss near edge structure (ELNES) probes the local bonding states in materials, we considered so-called nanoarcs, which consist of curved graphite sheets which are repeated in space either as wavy objects or as bonded arcs. For feasible calculations, we 
consider the extreme cases of the armchair (n,n) and the zigzag (n,0) nanotubes which form 
chiral angles of 30$^\circ$ and 0$^\circ$, respectively. All other nanotubes (chiral) 
have their chiral angles within the range [0$^\circ$-30$^\circ$]. 

This article is organised as follows. Section 2 is devoted to the description of the theoretical background. In section 3 we give the computational details. Section 4 is devoted to a discussion of the main results and we end in  section 5 with a brief summary.

\section{Theoretical framework}
\subsection{The augmented plane-wave method}
Within the density functional theory the ground state properties are derived from the ground state density $n({\bf r})$ which minimises the total energy functional
\begin{equation}
E_{tot}[n]=T[n]+U[n]+E_{xc}[n],
\end{equation}
where $T$ is the kinetic energy functional of the non-interacting electron system, $U$ is the electrostatic energy functional and $E_{xc}$ is that of the exchange and correlation energy. The electron density which minimises $E_{tot}[n]$ is obtained by solving self-consistently the Kohn-Sham equations~\cite{hohenberg64,kohn65} 
\begin{equation}
H\psi_i({\bf r})=\left[\hat{T}+V_{eff}({\bf r})\right]\psi_i({\bf r})=\epsilon_i\psi_i({\bf r})
\end{equation}
where  $\epsilon_i$ and $\psi_i$ are the Kohn-Sham eigenvalues and eigenfunctions, respectively. The effective potential $V_{eff}$ created by the electronic and nuclear charges is given in atomic units as
\begin{equation}
V_{eff}({\bf r})=\int d^3{\bf r}^\prime{n({\bf r}^\prime)\over |{\bf r}-{\bf r}^\prime|}-\sum_I{Z_I\over|{\bf r}-{\bf R}|}+V_{xc}({\bf r}).
\end{equation}
$Z_I$ is the charge of the $I$th nucleus located at ${\bf R}$.
The electron charge density is constructed from the Kohn-Sham eigenfunctions $\psi_i$ as
\begin{equation}
n({\bf r})=\sum_{i=1}^\infty f_i\mid\psi_i(\bf r)\mid^2
\end{equation}
with $f_i$ being the occupation number of the eigenstate $\psi_i({\bf r})$.

Many functionals exist for the exchange and correlation 
energy, the simplest and widely used being the local spin density 
approximation (LSDA) which exists in several forms~
\cite{hedin71,muruzzi78,perdew92}. Improvements on the locality 
are found by using the generalised gradient approximation (GGA) (e.g. Perdew et al.~\cite{perdetal92,perdew96}). 

The index $i$ in the basis set $\left\{\psi_i({\bf r})\right\}$ is easily 
understood in the language of the one-particle Bloch states $|\nu {\bf k}\left>\right.$ which are expanded 
into atomic-like waves inside of spheres centred  on atoms  (muffin-tin spheres) matched to plane waves in the interstitial region. The band state inside of sphere $s$ is written in the Slater's augmented plane-wave method as
\begin{equation}
\psi^s_{\nu{\bf k}}({\bf r})=\sum_{l=0}^\infty\sum_{m=-l}^lD^s_{lm}(\nu{\bf k})u_l(\epsilon_{\nu{\bf k}},r)Y^l_m({\hat{\bf r}})
\label{eq:augment}
\end{equation}   
where ${\bf \hat{r}}={\bf r}/r$ and the radial function $u_l(\epsilon_{\nu{\bf k}},r)$ is the solution of the radial Kohn-Sham equation in the spherically averaged crystal potential. The expansion coefficients $D^s_{lm}(\nu{\bf k})$ are determined from the boundary conditions at the sphere's surface. The normalisation condition on  $\psi^s_{\nu{\bf k}}({\bf r})$ may be written as
\begin{equation}
\sum_{l=0}^\infty\sum_{m=-l}^l\mid D^s_{lm}(\nu{\bf k})\mid^2=1,
\end{equation}
which is a sum over all the local partial charges associated with the band state $|\left.\nu{\bf k}\right>$. 
\subsection{Angular momentum projected local density of states and electron energy loss spectroscopy calculations}
In the above formulation the angular momentum projected density of states is written as
\begin{equation}
\chi^s_{l,m}(E)=\sum_{\nu{\bf k} }|D^s_{lm}(\nu{\bf k})|^2\delta\left(E-\epsilon_{\nu{\bf k}}\right).
\end{equation}
 The vector analogy of the p orbitals permits the local p density of states to be projected onto a cartesian coordinate system as~\cite{nelhiebel99} 
  \begin{equation}
\chi^s_{p_x,p_y,p_z}(E)=\sum_{\nu{\bf k} }|D^s_{x,y,z}(\nu{\bf k})|^2\delta\left(E-\epsilon_{\nu{\bf k}}\right),
\label{eq:dos-r}
\end{equation}
with $D^s_x=-{1\over \sqrt{2}}\left[D^s_{1,+1}-D^s_{1,-1}\right]$, $D^s_y=-{i\over \sqrt{2}}\left[D^s_{1,+1}+D^s_{1,-1}\right]$ and $D^s_z=D^s_{1,0}$.

We now introduce the expression central to the theory of electron scattering by the atoms of a crystal. The double differential scattering cross-section for the excitation of an atom by fast electrons is given in the first Born approximation by~\cite{inokuti71} 
\begin{equation}
{\partial^2\sigma\over \partial E \partial \Omega}={4\gamma^2\over a_0^2}{k\over k_0}{1\over Q^4}S({\bf Q},E),
\end{equation}
where $a_0$ is the Bohr radius, $\gamma=(1-\beta^2)^{-1/2}$ is the relativistic factor, ${\bf k_0}$ and ${\bf k}$ the fast electron wave vectors before and after interaction, respectively, and ${\bf Q}={\bf k_0}-{\bf k}$ is the momentum transfer. The dynamic form factor (DFF) $S({\bf Q},E)$ is defined as
 \begin{equation}
S({\bf Q},E)=\sum_{i,f}|\left<i|e^{i{\bf Q}\cdot{\bf r}}|f\right>|^2\delta\left(E+E_i-E_f\right)
\end{equation}
for excitations from states $\left.|i\right>$ with eigenvalue $E_i$ to state $\left.|f\right>$ with eigenvalue $E_f$.
In terms of the coefficients of the augmented plane wave expansion Eq.(\ref{eq:augment}), Nelhiebel et al.~\cite{nelhiebel99} explicitly wrote down the DFF  for scattering from an initial state $|nlm\left>\right.$ to the projection of the final states onto the angular momentum state $|l^\prime m^\prime\left>\right.$ as
\begin{eqnarray}
S({\bf Q},E)&=&2\sum_{l^\prime m^\prime}\sum_{L^\prime M^\prime}\sum_{\lambda \mu}\sum_{\lambda^\prime \mu^\prime}4\pi(-1)^{l^\prime+L^\prime}i^{\lambda-\lambda^\prime}\nonumber \\
& & \times(2l+1)\sqrt{\left(2\lambda +1\right)\left(2\lambda^\prime +1\right)\left(2l^\prime +1\right)\left(2L^\prime +1\right)}\nonumber\\
& &\times Y^\lambda_\mu({\hat{\bf Q}})^*Y^{\lambda^\prime}_{\mu^\prime}({\hat{\bf Q}})\left<j_\lambda(Q)\right>_{n\epsilon^\prime ll^\prime}\left<j_{\lambda^\prime}(Q)\right>_{n\epsilon^\prime lL^\prime}\nonumber\\
& &\times \left(\begin{array}{clcr}l&\lambda&l^\prime\\
0&0&0\end{array}\right)\left(\begin{array}{clcr}l&\lambda^\prime&L^\prime\\
0&0&0\end{array}\right)\nonumber\\
& &\times \sum_m\left(\begin{array}{clcr}l&\lambda&l^\prime\\
-m&\mu&m^\prime\end{array}\right)\left(\begin{array}{clcr}l&\lambda^\prime&L^\prime\\-m&\mu^\prime&M^\prime\end{array}\right)\nonumber\\
& &\times \sum_{\nu{\bf k}}D^s_{l^\prime m^\prime}(\nu{\bf k})D^s_{L^\prime M^\prime}(\nu{\bf k})^*\delta\left(\epsilon^\prime-\epsilon_{\nu{\bf k}}\right)
\label{eq:dff}\end{eqnarray}
with $\epsilon^\prime=E+\epsilon_{nl}$. The radial integrals $\left<j_\lambda(Q)\right>_{n\epsilon^\prime ll^\prime}$ are defined as
\begin{equation}
\left<j_\lambda(Q)\right>_{n\epsilon^\prime ll^\prime}=\int _0^{R_t}dR R^2 u_{nl}(R)j_\lambda(QR)u_{l^\prime}(\epsilon^\prime,R)
\end{equation}
where $u_{nl}(R)$ is the radial part of the core electron's wave function, $R_t$ is the radius of the muffin-tin sphere and $j_\lambda(x)$ is the spherical Bessel function of order $\lambda$.
In the small $Q$ limit the first two 3j symbols $\left(\begin{array}{clcr}\cdot&\cdot&\cdot\\
0&0&0\end{array}\right)$ boil down to the dipole selection rule $l^\prime=l\pm1,L^\prime=l\pm1$. For finite $Q$, Eq.(\ref{eq:dff}) incorporates the orientation resolved character of the ELNES and it can therefore be of interest in probing the level of anisotropy present in a crystal and its dependence on the microscope setting.  The cross terms in the DFF are those corresponding to the coupling of different final angular momenta ($l^\prime\ne L^\prime$) resulting from the coherent composition of the band state $|\left.\nu{\bf k}\right>$. These terms become important for large scattering angles. 
 
For nanotubes, anisotropy effects on ELNES will be studied with respect to the angle $\gamma$ between the electron beam and the normal to the walls of the tube. In all the orientation resolved ELNES calculations, the electron beam, the tube axis and the normal to the walls of the tube will all lie in the same plane. The microscope setting will be characterised by the beam convergence semi-angle $\alpha$ and the aperture collection semi-angle $\beta$. The dynamic form factor is calculated by integrating over all possible orientations of the momentum transfer within the solid angles subtended by the convergence and collection semi-angles. Integrated ELNES (relevant for polycrystalline samples) will be obtained by averaging over all possible electron beam orientations.

\subsection{Relevant coordinate system of the nanotube}
\begin{figure}[H]
\begin{center}
\includegraphics[width=3in,height=3in,angle=0]{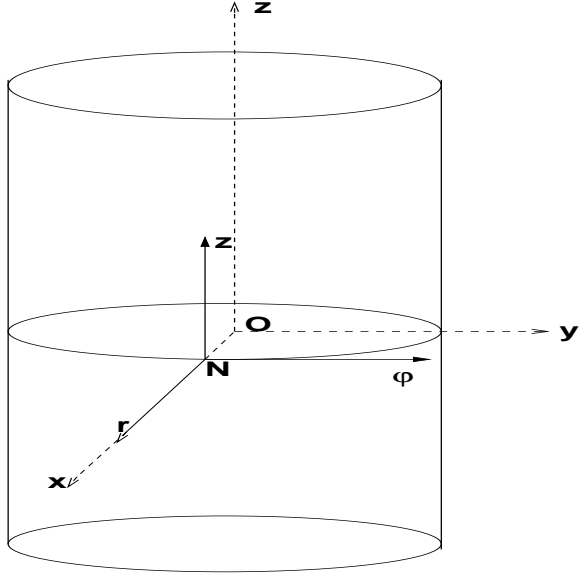}
\end{center}
  \caption{Relevant coordinate system of the cylindrical nanotube. We shall denote the ${\bf z}$ direction as axial, the azimuthal (${\bf \varphi}$) as perimetral and the normal to the walls of the tube as radial (${\bf r}$).} 
  \label{fig:tube-coord}
\end{figure}

On Figure~\ref{fig:tube-coord} we show the relevant coordinate system of the cylindrical nanotube. The decomposition of the local p density of states (pDOS) of Eq.(\ref{eq:dos-r}) for an atom sitting at the point ${\bf N}$ will be allocated as radial (for the r-pDOS), perimetral (for the $\varphi$-pDOS) and axial (for z-pDOS). All the coordinates of the atoms in the unit cell of a nanotube can be related by symmetry to those of a particular atom. Thus, the local electronic properties of all the atoms of the homogeneous defectless nanotube should be identical and the total DOS  should be a scalar multiple of the local DOS of an atom. The decomposition of the local pDOS of an atom into radial, perimetral and axial components should therefore reflect that of the total electronic density of state of the nanotube. 
  
\subsection{The pyramidalisation angle $\mu$ of zigzag and armchair tubes}
\begin{figure}[H]
\begin{center}
\includegraphics[width=3.in,height=3.in,angle=0]{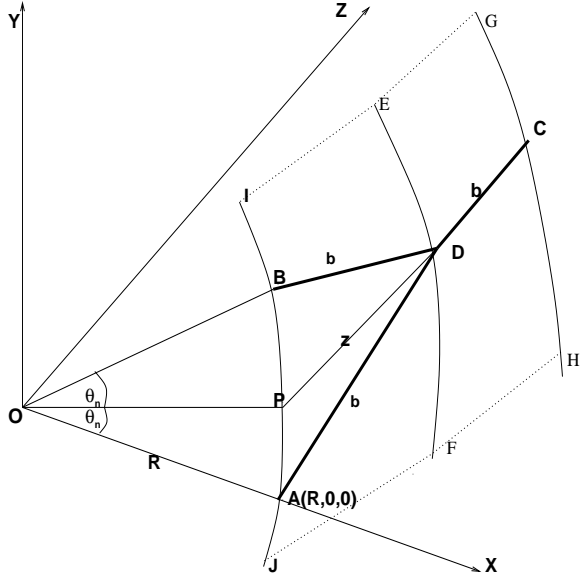}
\end{center}
  \caption{Geometry for determining the pyramidalisation angle of a zigzag (n,0) tube. OZ is along the tube axis and OX and OY are radial to the tube. IPJ, EDF and GCH are arcs of the tube.} 
  \label{fig:zigzag-pyramid}
\end{figure}

Curvature effects of the tubes on the properties can be studied with respect to the pyramidalisation angle $\mu$. To define the pyramidalisation angle, we consider the zigzag tube (n,0) whose arc is drawn on Figure~\ref{fig:zigzag-pyramid}. This arc is depicted by the arcs IPJ, EDF and GCH. OZ is along the tube axis. On the figure we show a carbon atom D bonded to three other atoms A, B and C. For the flat graphene sheet A, B, C and D all lie in the same plane but as curvature is introduced as a result of tube formation atom D becomes lifted out of the plane bearing A, B and C. In order to quantify the curvature of the nanotube on the local structure of a carbon atom, we may define the pyramidalisation angle $\mu$ for the 3-coordinated atom D as the angle between the normal to the plane bearing the three nearest neighbours (atoms A, B and C) and any of the 3 carbon bond vectors (bonded to D, either bond AD, BD or CD). The C$\large{-}$C bond length is fixed at b=1.42\AA~ and from Figure \ref{fig:zigzag-pyramid} it follows that $z=PD=\left(b^2-2R^2(1-cos\theta_n)\right)^{1/2}$ where R is the radius of the tube which for an ideal (n,m) tube is $R={\sqrt{3}b\over 2\pi}\left(n^2+m^2+nm\right)^{1/2}$.
Geometrical considerations lead to   
\begin{equation}
\mu_{(n,0)}-\pi/2=arc\sin\left[{\left(1-\cos(\theta_n)\right) \over\sqrt{\left({2\over\sqrt{3}}\theta_n\!+\!x_n\right)^2+\!\left(1-\!cos(\theta_n)\right)^2}}\right]
\end{equation}
where $x_n=\left(4\theta_n^2/3-2\left(1-\cos(\theta_n)\right)\right)^{1/2}$ and $\theta_n=\pi/n$. Similar considerations for the armchair tubes (n,n) result in
\begin{equation}
\mu_{(n,n)}-\pi/2={1\over 2}arc\sin\left[{\pi\over n}-\arccos\left(1-{1\over 2}\left({2\pi\over 3n}\right)^2\right)\right]\label{eqn:arm-mu}.
\end{equation}
To first order in $b/R$ we get $\mu\approx \pi/2+b/(4R)$ for both zigzag and armchair tubes which is the result obtained by Dumitric\u{a} et al.~\cite{dumitrica02}.
It is important to remark that if we define the $\pi$ orbital axis as one making equal angles with all three $\sigma$ bonds of a 3-coordinated atom then this angle will be equal to the pyramidalisation angle for the case of a nanotube. The deviation of this angle from 90$^\circ$ induces a new hybrid between the s and the p$_z$ (z in the language of the planar trigonal coordination) orbitals (s$^m$p$_z$). The proportion of s states $r=m/(m+1)$ in the $\pi$-orbital has been expressed in the $\pi$-orbital axis vector (POAV1) approximation~\cite{haddon88,haddon86a,haddon86b} as $r=2\cot^2\mu$ from which the new bonding state is sp$^{2+\delta}$ with $2+\delta=2(1-r)/(1-2r)$. In other words, a fraction $2r/(1-2r)$ of the p$_z$ orbital participates in the $\sigma$-bond. For the C$_{60}$ molecule $\delta\sim0.278$ meaning an s fraction of $r\sim0.11$ in the $\pi$ orbital of C$_{60}$. For tetrahedral coordination, $\mu=109.47^\circ$ corresponding to a sp$^3$ hybridisation.
 
\section{Computational details}
We study the electronic properties of a series of carbon nanotubes and some 
curved carbon systems using the ab initio all-electron 
Full-Potential-Linearised-Augmented-Plane-Wave (FPLAPW) 
package WIEN2k \cite{wien2k}. The unit cell is partitioned  
into two regions; regions of nonoverlapping spheres centred on the atoms and 
an interstitial region. The basis functions inside the muffin-tin spheres are 
linear combinations of spherical harmonics $Y^l_m({\hat{\bf r}})$ and radial functions $u_{l}(r)$. These are 
augmented with a plane wave basis in the interstitial region (with the 
coefficients being functions of the reciprocal lattice vectors and the wave 
vectors of the first Brillouin zone). The maximum $l$-value of 10 is adopted inside the atomic spheres. The charge density is Fourier expanded up to $G_{max}=14$. The Brillouin-zone integration is done using the Bl\"ochl~\cite{blochl94} 
improved tetrahedron method. The exchange and 
correlation energy is treated using the local spin density approximation~\cite{perdew92} which is a fit to the Green's-function Monte Carlo calculations of Ceperley and Alder~\cite{ceperley80}. Core states are treated fully relativistically.  
    
The choice of the muffin-tin radius and the plane wave energy cut-off are governed 
by the parameter $RKM={ R_{MT}\times K_{max}}$ (where $R_{MT}$ is the 
smallest muffin-tin radius in the crystal). For these carbon 
systems the 1s state is the core state while 2s and 2p are the 
valence states.  Muffin-tin radii of 1.3 a.u. are used. The total number of 
k-points in the full Brillouin zone was fixed at 1000 (between 60 and 80 in the irreducible wedge, depending on the dimensions 
of the unit cell) for the self consistency calculations without the core-hole effect and 300 (between 100 and 150 in the irreducible wedge - due to lower symmetry) for core-hole calculations in larger unit cells. Due to the supercell nature of 
each unit cell of the nanotube, these values should be representative of infinite samplings 
of the Brillouin zones. The tubes (and also the nonoarcs) are arrayed in orthorhombic unit cells. The ideal graphene folded cylinders are adopted for all the tubes studied. For the (6,6) tube the difference in DOS and ELNES for RKM values of 5, 6 and 7 were insignificant. So we adopted an RKM value of 5.5 for all other tubes.

When a core electron is excited into the unoccupied level a core-hole is left behind which interacts with the valence and conduction  electrons and modifies the crystal electronic structure if the core-hole is not well screened by the valence and conduction electrons. In metals the screening is effective meanwhile in semiconductors there is less screening and the core-hole interaction becomes very important. This core-hole interaction is a dynamical process and should be treated dynamically using the Green's-function approach~\cite{hedin65,hybertsen85}. Two ad hoc techniques are widely used to account for this core-state effect. These include the Z+1 method or the equivalent core approximation~\cite{ma93} whereby the excited core seen by the valence and conduction electrons is equivalent to that of an atom with atomic number Z+1. For the case of carbon systems the core atom is replaced by a nitrogen atom. The second method, which we term the core-excited method, consists of introducing a core-hole in the core state by removing an electron from it and inserting it as a uniform background charge. In the so-called ``sudden'' approximation no core-hole effect is considered and ELNES or any excitation process is calculated from the ground state electronic structure.

Eq.(\ref{eq:dff}) has been implemented into the WIEN2k package~\cite{wien2k}.
ELNES was calculated on the lowest symmetry-adapted k-mesh of each system where all symmetry operations are eliminated except the identity. By doing so cross terms of the dynamic form factor are well reproduced.  The core-hole effect is introduced in all the tubes studied via the core-excited method. For the armchair tubes (n,n) the unit cell was doubled along the tube axis in order to increase the distance between core atoms. This means a minimum distance between core atoms of 4.91\AA~and 8n atoms per unit cell. Since the length of the smallest unit cell along tube axis was 4.26\AA~ for the zigzag tube (n,0), these smallest unit cells with 4n atoms were used for all such tubes with one of the 4n atoms made the core atom. The minimum wall-to-wall distance was set at 4\AA~ to reduce intertube interactions. 

\section{Results and discussions}
The various tubes  studied together with their diameters and pyramidalisation angles are shown on Table~\ref{tabl:tubes}. In order to study the curvature effect on the electronic properties of nanotubes, we also considered the case of the (4,0) tube which has not been shown to exist in nature. The (5,0) tube has recently been synthesised in a zeolite cage~\cite{li01}.
\begin{table}[H]
\begin{center}
\begin{tabular}{|c|c|c||c|c|c||c|c|c|}
\hline
Tube& Diameter (\AA)&$\mu_n$($^\circ$)& Tube& Diameter (\AA)&$\mu_n$($^\circ$)&Tube& Diameter (\AA)&$\mu_n$($^\circ$)\\
\hline
(4,0)& 3.1& 101.9& (7,0)& 5.5&97.2& (4,4) &5.4 &97.4\\
\hline
(5,0)& 3.9&  99.8& (8,0)&        6.3 &       96.2 & (5,5) &        6.8 & 96.0\\
\hline
(6,0)& 4.7&  98.3&(10,0)&         7.8&        95.1& (6,6) &        8.1 & 95.0\\
\hline
\end{tabular}
\end{center}
\caption{The different tubes studied; their diameters and pyramidalisation angles.}
\label{tabl:tubes}
\end{table}

\subsection{Total and local density of states}
\begin{figure}[H]
\begin{center}
\includegraphics[width=4.in,height=6.in,angle=-90]{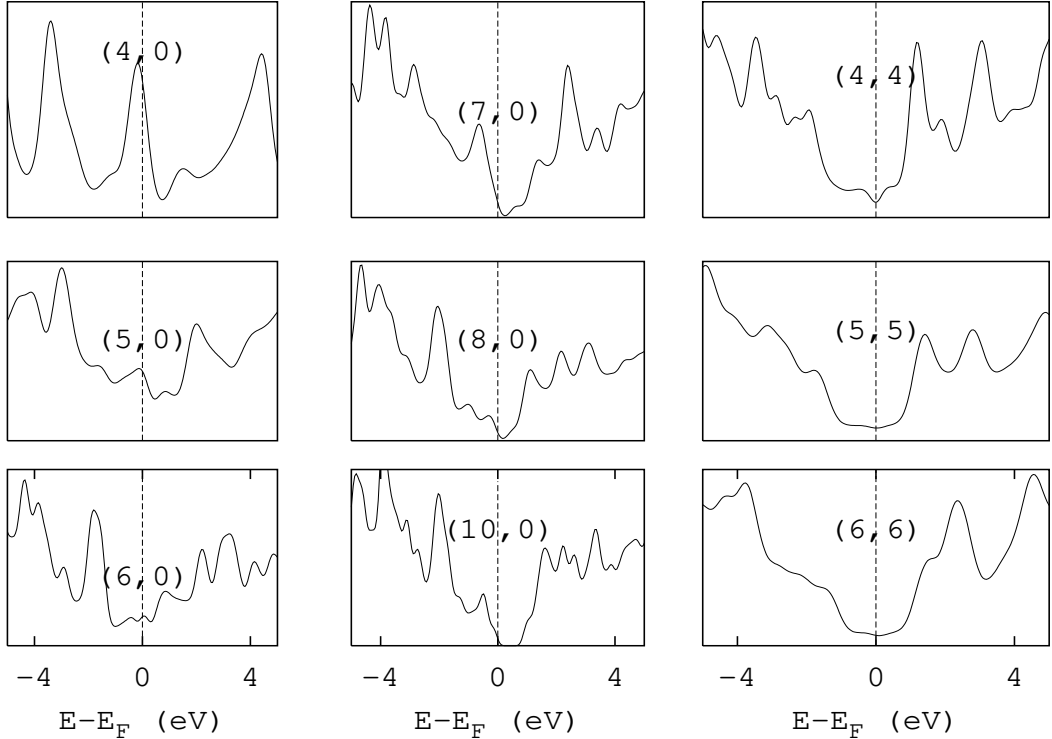}
\end{center}
  \caption{Total DOS of all the studied tubes. Spectra were broadened  using Gaussians of FWHM of 0.1 eV} 
  \label{fig:TDOS}
\end{figure}

In the zone-folding tight-binding approach, any nanotube (n,m) is metallic if 
$\mid n-m\mid$ is divisible by 3 \cite{saito98}. From this it follows that all 
armchair tubes are metallic. Figure~\ref{fig:TDOS} shows the ground state 
total DOS of all the  studied tubes.  

\begin{figure}[H]
\begin{center}
\includegraphics[width=4.0in,height=6.in,angle=-90]{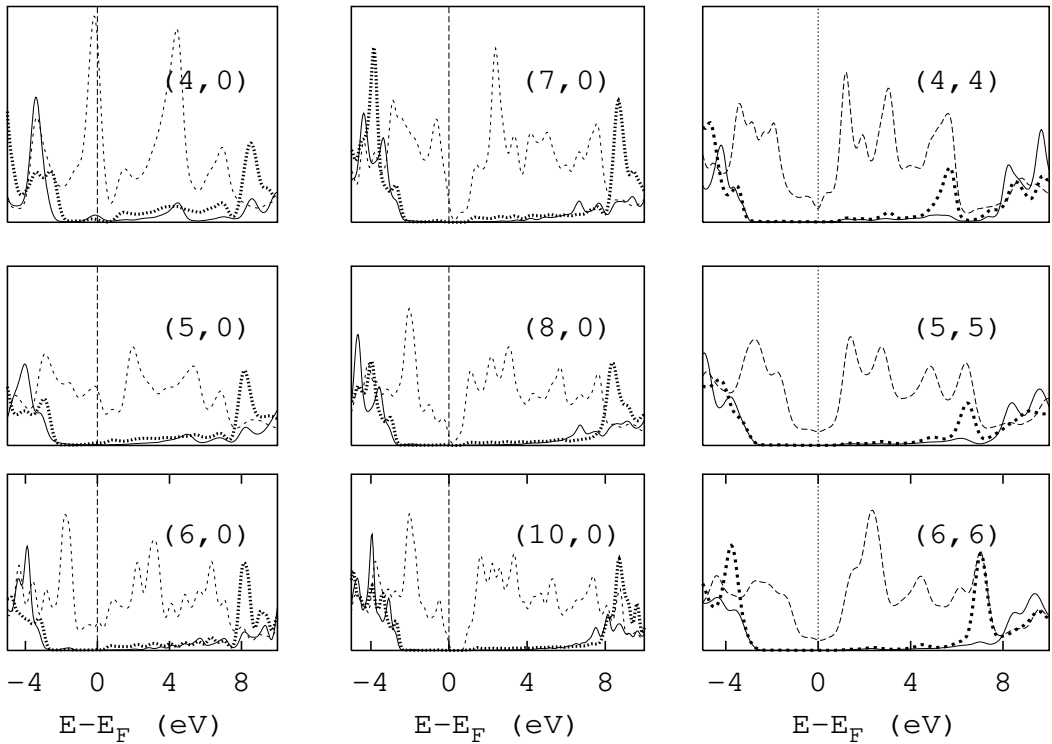}
\end{center}
  \caption{Axial (full line), radial (dashed line) and perimetral (dotted line) local DOS for nine nanotubes. The vertical line denotes the Fermi level.} 
  \label{fig:metallicity}
\end{figure}

Remark the feature at the Fermi level for the strongest curved tube (4,0). This feature is also present for all 
other small zigzag tubes but to lesser intensity as the curvature decreases. 
It vanishes for the (10,0) tube. Such a remarkable curvature dependent feature 
is absent for all armchair tubes whose DOS remains flat within $\pm 1$ eV 
around the Fermi level except for a depression of the DOS at 
the Fermi level for the small (4,4) tube.  Such a depression was also 
reported by Delaney et al.~\cite{delaney98} on a DFT calculation on the 
(10,10) tube and this was due to inter-tube interactions. The metallicity condition according to the zone-folding tight-binding theory is based on the $\pi$ and $\pi^*$ bands. The notion of $\pi$ and $\sigma$ becomes ill-defined for these curved systems but we may speak of curvature induced $\pi$-$\sigma$ ($\pi$ and $\sigma$ of the planar graphene) hybridisation to form new hybrids. A careful study of the local density of states (for a particular atom) in all three directions (axial, radial and perimetral to the tube) reveals some interesting aspects: in that while the finite LDOS at the Fermi level of all the small tubes are dominantly radial to the tube (see Figure~\ref{fig:metallicity}), non negligible axial and perimetral LDOS are present around the Fermi level for the small zigzag tubes. This is a signature of $\pi-\sigma$ hybridisation. This hybridisation is  weak for tubes whose diameters are greater than 5.5\AA~ irrespective of chirality. This $\pi-\sigma$ hybridisation may be quantified in the $\pi$-orbital axis vector (POAV1) approximation~\cite{haddon88,haddon86a,haddon86b} described above. In this approximation, as much as 22\% p$_z$ electrons participate in the $\sigma$ bonding for the (4,0) tube while less than 4\% participates for the (7,0) and (4,4) tubes of similar diameters~$\sim$5.5\AA. It is important to remark that while the curvature-induced hybridisation drastically modifies the electronics of small diameter zigzag tubes its effect on arm chair tubes is weaker.   
 Blase et al.~\cite{blase94} studied the hybridisation effects on the metallicity of small radius zig-zag NTs and concluded that the curvature induced $\pi-\sigma$ hybridisation effects change the energy and character of lowest lying conduction band states with important consequences on the metallicity and transport properties of the tubes. Reich and Thomsen~\cite{reich02} also pointed out that, compared to the zigzag tubes, the electronic band structure is less affected in the armchair and chiral tubes by the curvature induced $\pi-\sigma$ hybridisation.  
    
\subsection{Energy loss near edge spectroscopy (ELNES)} 
\subsubsection{Effects of core-hole interaction}

For the (4,0), (5,0), (7,0), (10,0), (5,5) and (6,6) tubes we performed core-hole-included DFT calculations, using the excited-core approach, by removing an electron from the 1s state and inserting it as a uniform background charge. For the (5,0) tube we also introduced the core-hole effect through the alternative Z+1 method~\cite{ma93} which essentially gave the same result.
These core-hole techniques have also been  applied on graphite and are found to yield the same results which are in good agreement with the ELNES measurements~\cite{graphite} up to about 15 eV beyond the threshold. 

\begin{figure}[H]
\begin{center}
\includegraphics[width=4.3in,height=5in,angle=-90]{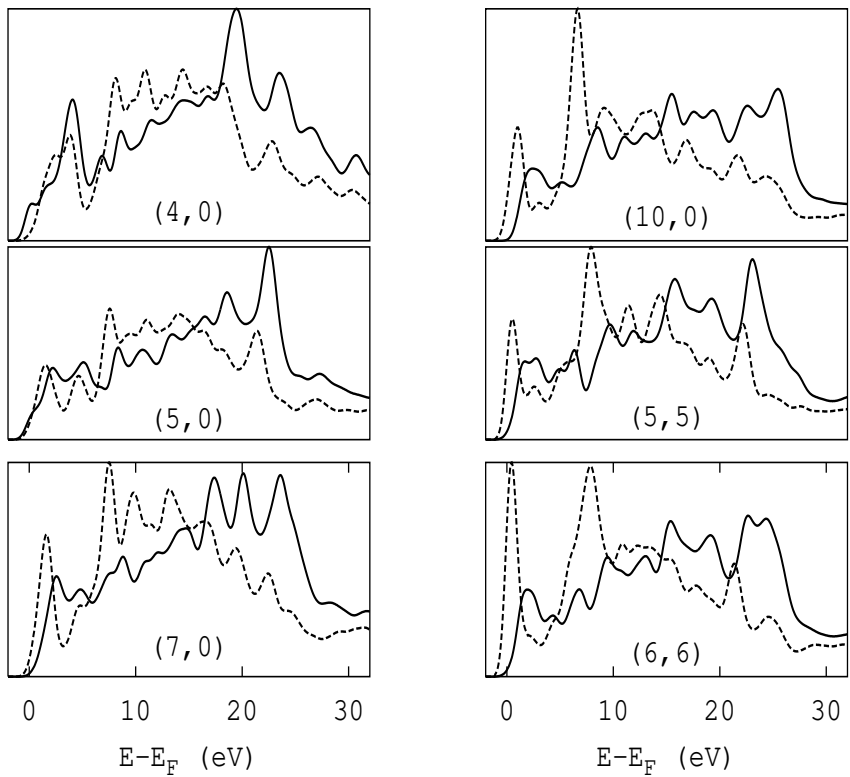}
\end{center}
  \caption{Comparison between the sudden approximation (full lines) and core-hole (dashed line) calculated ELNES of (4,0), (5,0), (7,0), (10,0), (5,5) and (6,6) tubes and the curvature effect on the ELNES of these tubes. The Fermi level is fixed as the energy origin.}    
  \label{fig:core-hole-NT}
\end{figure}

For the zigzag tubes the minimum distance between core atoms was set at 4.26 \AA~ while for the armchair tubes it was fixed at 4.91 \AA. Due to this difference in the inter-core-atom distances, we expect 
stronger intratube interactions between the core atoms of the zigzag tubes than in the armchair  tubes though this interaction will not strongly influence the C K-edges.  Intertube core-atoms interactions can be ignored as the least distance between such atoms is more than 7\AA. On Figure~\ref{fig:core-hole-NT} we compare the effects of the inclusion of core-hole on the C K-edge of six tubes. To account for instrumental and life-time broadening we convoluted the calculated spectra with Gaussians with full width at half maximum (FWHM) of 0.5 eV. The microscope convergence and collection semi-angles ($\alpha$ and $\beta$) were set at the magic values (1.87, 3.01 mrad). Orientation averaged spectra are shown though at this microscope setting no orientation effect should be seen. 

The immediate effect of the core-hole interaction is the shifting of the $\pi^*$ edge to lower energies and the redistribution of $\sigma^*$ intensity to the lower energy part. A signature of $\pi-\sigma$ hybridisation is the feature at 5 eV  which decreases in intensity as the tube diameter increases from the (4,0) tube through the (7,0) tube and is evidenced in the armchair tubes as the shoulders of  the lower energy part of the $\sigma$ peaks. Comparison with measurements on SWNT~\cite{knupfer99,stephan01,knupfer01} reveals that the core-hole must be considered if ELNES measurements are to be correctly explained theoretically.

Knupfer et al.~\cite{knupfer99} detected a new feature in a single-wall carbon nanotube 1s EELS spectrum about 2 eV above the $\pi^*$  peak (i.e. about 5 eV above the Fermi level), which was absent in some samples~\cite{knupfer01}. The origin of this feature remained unexplained. On Figure~\ref{fig:core-hole-NT} we remark the $\pi-\sigma$ hybridisation features 2-3 eV above the main $\pi^*$  peak of small tubes. This may suggest that in Knupfer's sample there could have been tubes as small as 4-10 \AA~ in diameter.  

\begin{figure}[H]
\begin{center}
\includegraphics[width=4.in,height=5in,angle=-90]{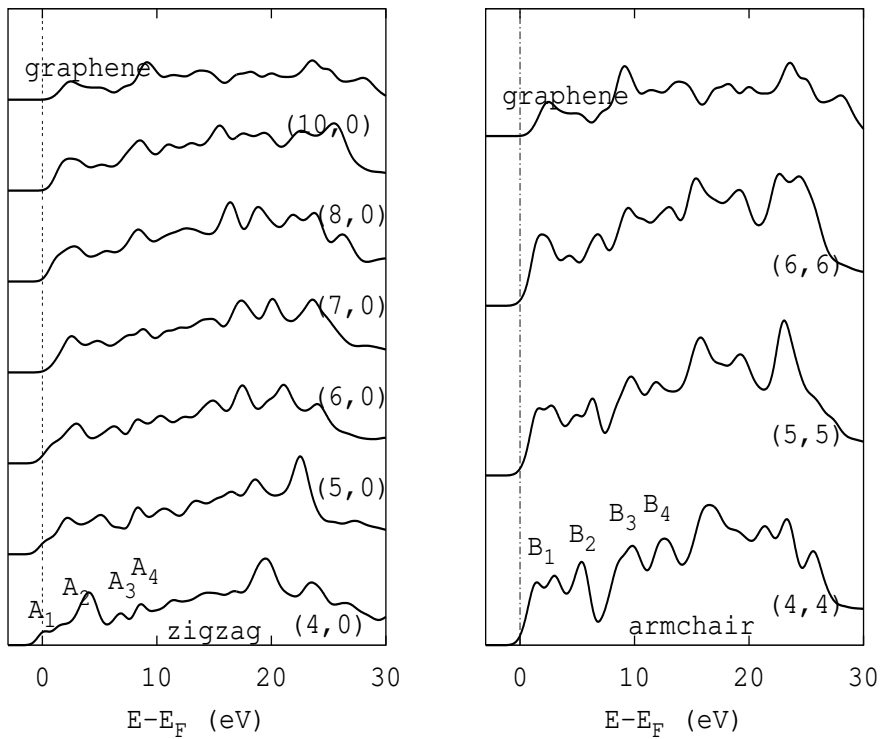}
\end{center}
  \caption{Effects of curvature  on the ELNES of small armchair and zigzag tubes. No core-hole calculation is implemented. Each spectrum is shifted upward for proper visualisation. The vertical lines denote the Fermi level.}    
  \label{fig:elnes-cuvature}
\end{figure}

ELNES spectra were also calculated in the so-called ``sudden'' approximation for all the nine tubes and compared with that of flat graphene in Figure~\ref{fig:elnes-cuvature}. Although it is very clear by now that ELNES calculations on nanotubes that do not include the core-hole effects cannot be used to explain experimental EELS measurements, such calculations should lead to insights in 
the ground state momentum projected local density of states and the effects of
 curvature or local deviations from planar configurations on the transition 
matrix elements. As Figure~\ref{fig:elnes-cuvature} shows, the direct effect 
of curvature is the splitting of the $\pi^*$ peak into the A$_1$ and A$_2$ 
(B$_1$ and B$_2$) features for the zigzag (armchair) tubes and the splitting 
of the $\sigma^*$ peak into the A$_3$ and A$_4$ (B$_3$ and B$_4$) features 
for the zigzag (armchair) tubes. Suenaga et al.~\cite{suenaga01} performed 
the ELNES calculation of (5,0) and (10,0) tubes (in the sudden approximation) 
and reported the splitting of the $\pi^*$ peak as curvature increases. They 
also pointed out that as the curvature increases the $\sigma^*$ peak 
decreases. We remark also that as the curvature increases the predominantly 
$\pi^*$ peak A$_1$ diminishes while the intensity of the hybrid feature A$_2$ 
grows. The weak overlap between B$_2$
 and  B$_3$ and the strong overlap between A$_2$ and A$_3$ for the strongest curved armchair tube (4,4) and strongest curved zigzag tube (4,0), respectively, demonstrate that B$_2$ is purely of $\pi^*$ character, B$_3$ purely of $\sigma^*$  character while A$_2$ and A$_3$ are of mixed $\pi^*$ and $\sigma^*$ characters. This point emphasises the fact that the curvature induced $\pi-\sigma$ hybridisation effect is stronger in zigzag tubes than in armchair tubes.  From Figure~\ref{fig:core-hole-NT} we also notice that the core-hole leads to a clear distinction between the $\pi^*$ and the $\sigma^*$ edge onsets. 

\subsubsection{Energy dependence of the transition matrix elements}
\begin{figure}[H]
\begin{center}
\includegraphics[width=3.2in,height=3.5in,angle=-90]{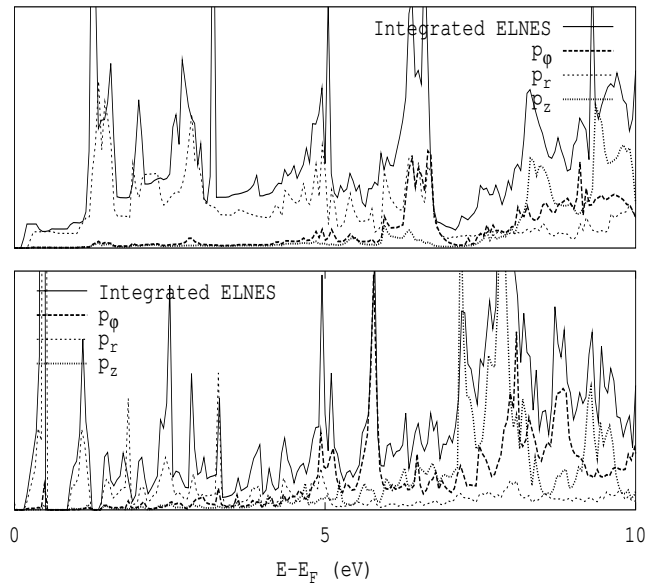}
\end{center}
  \caption{The perimetral LDOS (p$_{\varphi}$), the radial LDOS (p$_{r}$) and axial  LDOS (p$_{z}$) of a carbon atom of the (5,5) 
nanotube and the ELNES of the same atom. The upper panel shows the non core-hole-included 
results while the lower panel shows the effect of core-hole.} 
  \label{fig:matrix-elt}
\end{figure}

\begin{figure}[H]
\begin{center}
\includegraphics[width=3.5in,height=3.5in,angle=-90]{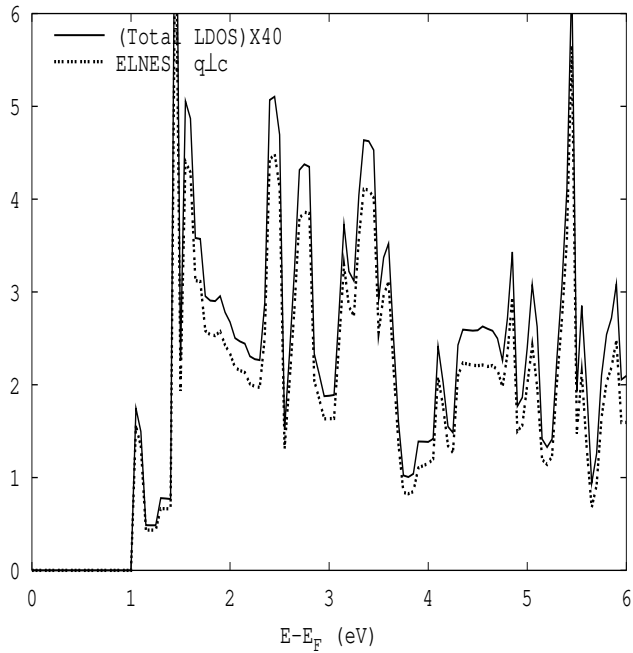}
\end{center}
  \caption{Local density of states (LDOS) of the carbon atom of the (10,0) 
nanotube and the ELNES of the same atom with core-hole included: momentum transfer orientation resolved calculations is shown.} 
  \label{fig:matrix-elt-10-0}
\end{figure}

In order to investigate the relationship between angular momentum projected local DOS and the 
calculated ELNES and the effect of core-hole on the transition matrix elements we plot the radial
(p$_{r}$), perimetral (p$_{\varphi}$) and axial (p$_{z}$) p density of states of an atom together with the unbroadened ELNES for the (5,5) tube on 
Figure~\ref{fig:matrix-elt}. The orientation averaged ELNES spectra are calculated for microscope convergence and collection semi-angles of 1.87 and 3.01 mrad, respectively.  
The calculated ELNES features are excellently reproduced by the projected DOS for both non core-hole and core-hole calculations. The spikes in the DOS are also beautifully reproduced 
by both ELNES spectra. Figure~\ref{fig:matrix-elt-10-0} now shows the case of a (10,0) tube. Remark that the agreement between LDOS and ELNES is excellent for the momentum resolved ELNES. This strong agreement between ELNES features and LDOS shows the weak energy dependence of the transition matrix elements though these will be shown to depend strongly on the electron beam orientation and the microscope settings. It also shows that the dipole selection rule governs the ELNES spectrum for nanotubes because all the ELNES features are those of the pLDOS. Knupfer et al.~\cite{knupfer99} reported the ELNES spectrum of a SWNT which was free of singularities and concluded that the core-hole effect was responsible for this. Though their measurements were done using a high energy resolution microscope (0.115 eV), which should be able to detect the spikes, these were conspicuous by their absence. Thermal broadening could have been responsible for the washing out of these singularities.

\subsubsection{Orientation resolved ELNES}
On Figure~\ref{fig:orientation-6,6_10-0} we plot the total ELNES of (6,6) and (10,0) tubes and graphite for two incident electron beam orientations. The electron beam makes an angle $\gamma$ with the normal to the graphite plane or the walls of the nanotubes. When $\gamma=0^\circ$ momentum transfer is parallel to the walls of the tube and scattering is mostly due to the in-plane bonds. For $\gamma=90^\circ$ momentum transfer is normal to the walls of the tube and scattering probes off-plane bondings. Both the (6,6) and (10,0) tubes spectra are for core-hole calculations and are compared with core-hole calculated graphite. 
Orientation resolved ELNES calculations for the (6,6) and (10,0) tubes reveal that the magic convergence and collection couple ($\alpha^*$,$\beta^*$)=(1.87,3.01mrad) found for 
graphite remains magic for these curved carbon systems (and possibly for all other carbon materials) as Figure~\ref{fig:orientation-6,6_10-0} shows. Anisotropy is seen for $\beta\ne\beta^*$ when $\alpha^*$ is fixed at 1.87 mrad. For $\beta<\beta^*$ 1s$\rightarrow\sigma^*$ transition is strong for the electron beam perpendicular to the graphene plane ($\gamma=90^\circ$). This tendency reverses for $\beta>\beta^*$, all in similar trend as in graphite.  St\'ephan and co-workers~\cite{stephan01} measured the ELNES for the electron beam normal ($\gamma=0^\circ$)  and perimetral ($\gamma=90^\circ$) to a multi-wall nanotube and found that the anisotropy was weak compared to that of graphite. A look at the topmost two panels of Figure~\ref{fig:orientation-6,6_10-0} confirms this finding. Following the discussion of the previous section, we can conclude that this anisotropy is entirely governed by the transition matrix elements  and that the latter depend strongly on the electron beam characteristics and the microscope settings in a manner well elaborated in refs.~\cite{menon98,menon99}.

\begin{figure}[H]
\begin{center}
\includegraphics[width=3.5in,height=6in,angle=-90]{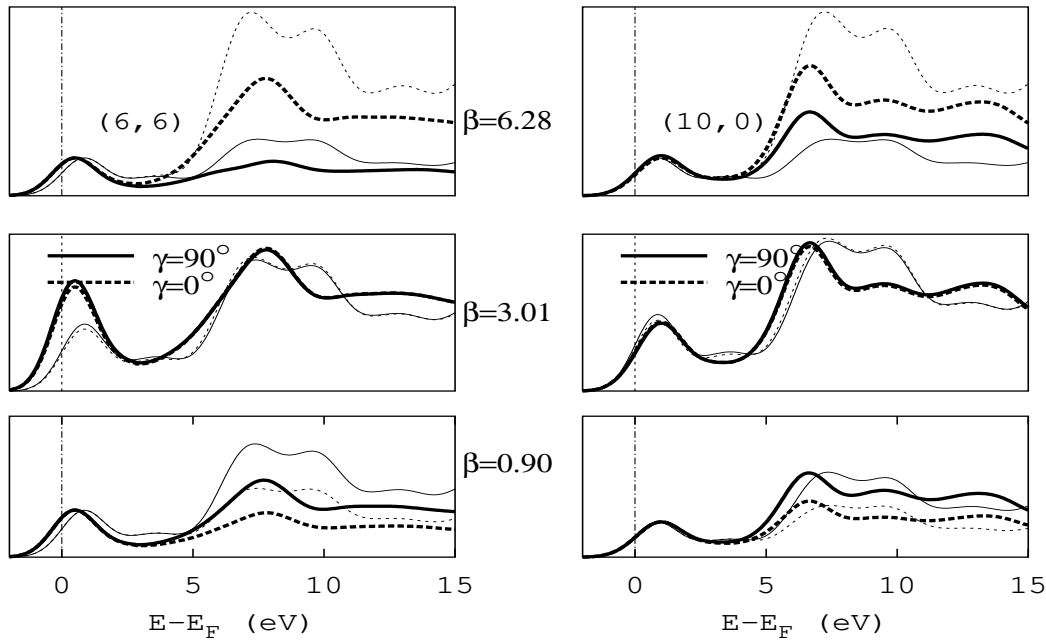}
\end{center}
  \caption{Orientation resolved ELNES of (6,6) and (10,0) tubes (bold line) and comparison with graphite (thin line). The spectra for $\gamma=0^\circ$ (dashed line) and $\gamma=90^\circ$ (full line) coincide exactly over all energies for $\beta=3.01$ mrad but no such coincidence is seen for $\beta=6.28$ mrad and $\beta=0.90$ mrad. The vertical line indicates the Fermi level.} 
  \label{fig:orientation-6,6_10-0}
\end{figure}
\begin{figure}[H]
\begin{center}
\includegraphics[width=3.5in,height=5in,angle=-90]{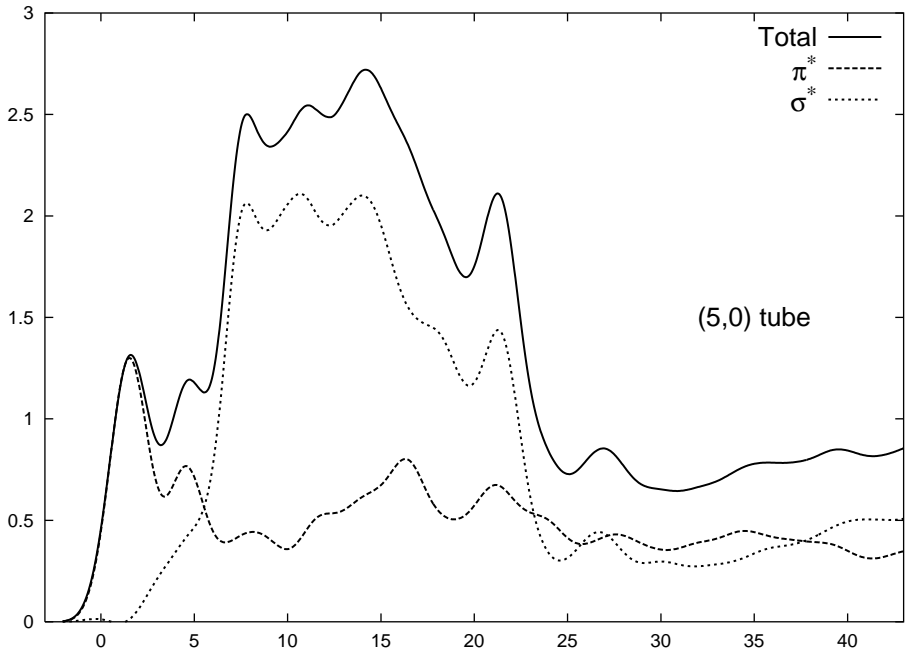}
\end{center}
  \caption{Spectral (l$^\prime$,m$^\prime$) decomposition of the ELNES spectrum of the (5,0) tube. The total spectrum (full line) for $\gamma=0^\circ$ are decomposed into the $\pi^*$  and the $\sigma^*$ components. Remark that as a signature of $\pi-\sigma$ hybridisation, the tail of the $\sigma^*$ component overlaps with the $\pi^*$ spectrum. A Gaussian broadening of 1.8 eV FWHM is used.} 
  \label{fig:orientation-5-0}
\end{figure}

We performed a ($l^\prime$,$m^\prime$) decomposition~\cite{nelhiebel99} (which consists of the various terms of Eq.(\ref{eq:dff}) when $l^\prime$ and $m^\prime$ are fixed) of the ELNES in a fixed reference frame. In this decomposition, the 
1s$\rightarrow 2p$ ELNES has three components: (1,0), (1,-1) and (1,+1) components. In the case of a planar system where the z-axis is perpendicular to the plane (for example, graphite) the 
(1,0) component will correspond to the 1s$\rightarrow \pi^*$ transition while (1,-1)+(1,+1) will correspond to the 1s$\rightarrow \sigma^*$ transition.  In the case of the (6,6) and (10,0) nanotubes, where the z-axis was  along the tube axis, the (1,0) component becomes a transition into the axial $\sigma^*$ orbitals. Unfortunately the (1,-1) and (1,+1) components become inseparable  mixtures of $\pi^*$ and the perimetral $\sigma^*$ components for these cases (see Figure~\ref{fig:orientation-10,0}). It is important to point out that the ELNES spectra on Figure~\ref{fig:orientation-10,0} are entirely given by the three p ELNES-components for all the three collection angles. This illustrates the fact that even for relatively large collection angles, the dipole sellection rule is still valid. In the case of the (5,0) tube we fixed the z-axis normal to the walls of the tube such that the (1,0) component can be obtained separately. On Figure~\ref{fig:orientation-5-0} we show the total ELNES, the $\pi^*$ and the $\sigma^*$ components of the core-hole calculated ELNES of the (5,0) tube. Remark that, as a signature of $\pi-\sigma$ hybridisation, the tail of the $\sigma^*$ component overlaps with the $\pi^*$ spectrum. This is in agreement with the observation (see Figure~\ref{fig:metallicity}) that around the Fermi level there is a non negligible DOS resulting from perimetral and axial $\sigma$ orbitals for the (5,0) tube. According to the POAV1 approximation~\cite{haddon88,haddon86a,haddon86b}, 14\% of the p$_z$ orbitals should participate in the hybrid orbital formation for the (5,0) tube. This agrees with the 15\% obtained when two Gaussian centred functions, one (P$_1$) describing the overlapping lower energy tail of the $\sigma^*$ edge and the other (P$_2$) for the main $\sigma^*$ peak, are used to fit the $\sigma^*$ edge as shown on Figure~\ref{fig:orientation-5-0}.  The 15\% was given by the ratio Area(P$_1$)/Area(P$_1$+P$_2$). Orientation resolved ELNES measurements should be able to reveal this rehybridisation in such small tubes. This technique of (l$^\prime$,m$^\prime$) decomposition has recently been applied to hexagonal BN~\cite{hebert00} and graphite~\cite{graphite}, both of which have planar symmetry. We believe that this $\pi^*$/$\sigma^*$ decomposition in planar systems is a viable alternative to the existing techniques~\cite{fallon93, bruley95,papworth00,berger88} for sp$^2$/sp$^3$ quantification of carbon materials based on the isolation of the $\pi^*$/$\sigma^*$ excitation cross-sections. 

 \begin{figure}[H]
\begin{center}
\includegraphics[width=3.5in,height=5in,angle=-90]{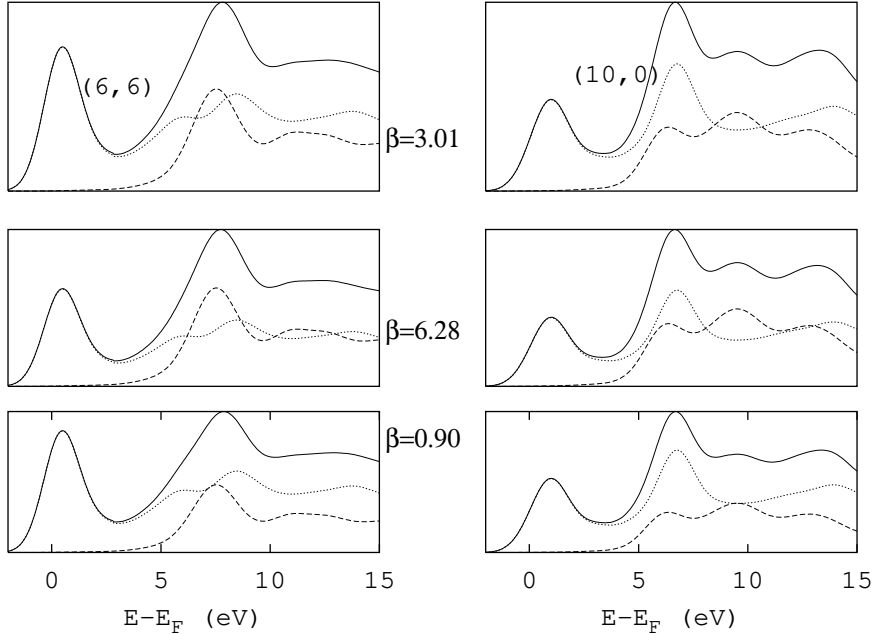}
\end{center}
  \caption{Spectral (l$^\prime$,m$^\prime$) decomposition of the ELNES spectra of the (6,6) and (10,0) tubes. The total spectrum (full line) for $\gamma=0^\circ$ are decomposed into the (1,0) or axial component (dashed line) and the (1,-1)+(1,+1) or radial+perimetral terms (dotted line). } 
  \label{fig:orientation-10,0}
\end{figure}
\subsection{Investigation of the local nature of ELNES}
It is usually postulated that ELNES probes the local properties  of materials. In particular, spatial and momentum resolved ELNES are often used as fingerprints of bonding states and local disorder or to explore the presence of grain-boundaries in materials. Due to the strong 
curvature applied to graphene to form a nanotube, carbon can undergo a wide range of bonding states from sp$^2$ for flat graphene to sp$^{2.28}$ for tubes with similar diameters as the C$_{60}$ molecule. This may imply a curvature induced re-activity of nanotubes different from that of graphite~\cite{menon00}.

In this section we consider arcs of tubes which are repeated in space periodically as bonded or wavy arcs in order to investigate the local nature of the energy loss near edge structures. We calculated the DOS and ELNES of these arcs and compared them with those of the full tubes.  
St\'ephan et al.~\cite{stephan01} simulated the curvature effects on the 
unoccupied density of states of NT's by considering a corrugated graphene sheet in which the corrugation was defined by the chair configuration of the carbon hexagons with a corrugation angle of 8$^\circ$ (an atom is lifted out of the plane such that the lifted bond vectors make an angle of 8$^\circ$ with the plane) to model a 10 \AA~ tube. Such a model which considers only the corrugation angle and ignores the chiral angle cannot be expected to be representative (electronically) of tubes as small as 10 \AA~ in diameter. Our bonded/wavy arcs are aimed at addressing this problem.

\begin{figure}[H]
\begin{center}
\includegraphics[width=2.5in,height=6in,angle=-90]{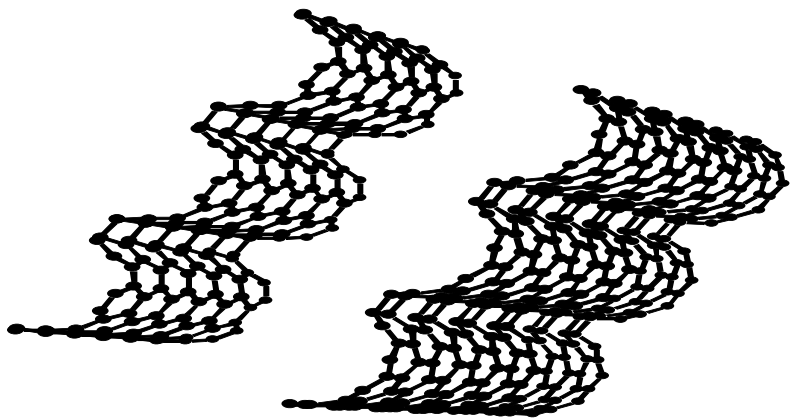}
\end{center}
  \caption{Armchair and zigzag bonded arcs.} 
  \label{fig:bonded}
\end{figure}

\begin{figure}[H]
\begin{center}
\includegraphics[width=2.5in,height=6in,angle=-90]{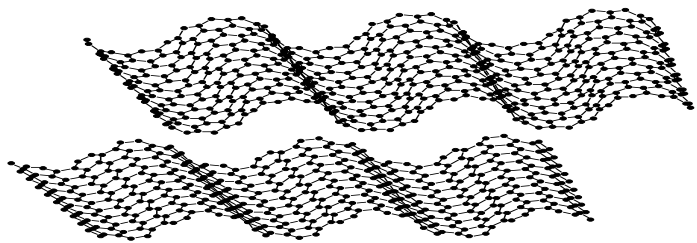}
\end{center}
  \caption{Zigzag and armchair wavy arcs.} 
  \label{fig:wavy}
\end{figure}

We define arcs of a given nanotube, e.g. an arc of an armchair tube (n,n) as a 
section of the tube that is cut out and is repeated in space (because of the use of the periodic boundary conditions) either by bonds (of the order of C-C 
bond length - see Figure~\ref{fig:bonded}) or as the wavy objects
shown on Figure~\ref{fig:wavy}. This latter model diminishes the unphysical 
hybridisations that are introduced by the boat configurations (for armchair arcs) and the 8 atoms ring configurations (for zigzag arcs) in the bonded arcs. The size of each arc is chosen large enough so that at least one of its carbon atoms is situated at a minimum distance of two C$\Large{-}$C bond-lengths from the point of contact of two arcs. This is to ensure a reduced interaction between the defectuous points and the carbon atom whose local properties should be comparable to those of an atom in a defect-free tube of similar curvature and chirality.

On Figures~\ref{fig:dos-eels-6,6} and 
\ref{fig:dos-eels-10,0} we show the total 
density of states and the ELNES spectra of the  C K-edges of the (6,6) 
and (10,0) tubes, respectively. For the ELNES spectra on these figures no core-hole 
calculations are used. The DOS and the carbon 1s excitation edges are also shown for 
the nanoarcs. The wavy and the bonded arcs reproduce 
qualitatively well (up to 12 eV beyond the edge onset) the ELNES of the (6,6) and the (10,0) tubes. These good agreements of the calculated ELNES of curved arcs with those of tubes of similar radius of curvature and chirality confirm the fact that ELNES (up to 10-15 eV above threshold) probes local structures. The total density of states, on the other hand, is a property 
that depends on the underlying periodicity. It should be strongly affected by 
periodically repeated defects like the boats and 8 atoms rings present in the bonded arcs. While the LDOS describes the local electronic structures, the total DOS is responsible for the observed macroscopic properties like metallicity. 

 \begin{figure}[H]
\begin{center}
\includegraphics[width=3.3in,height=6in,angle=-90]{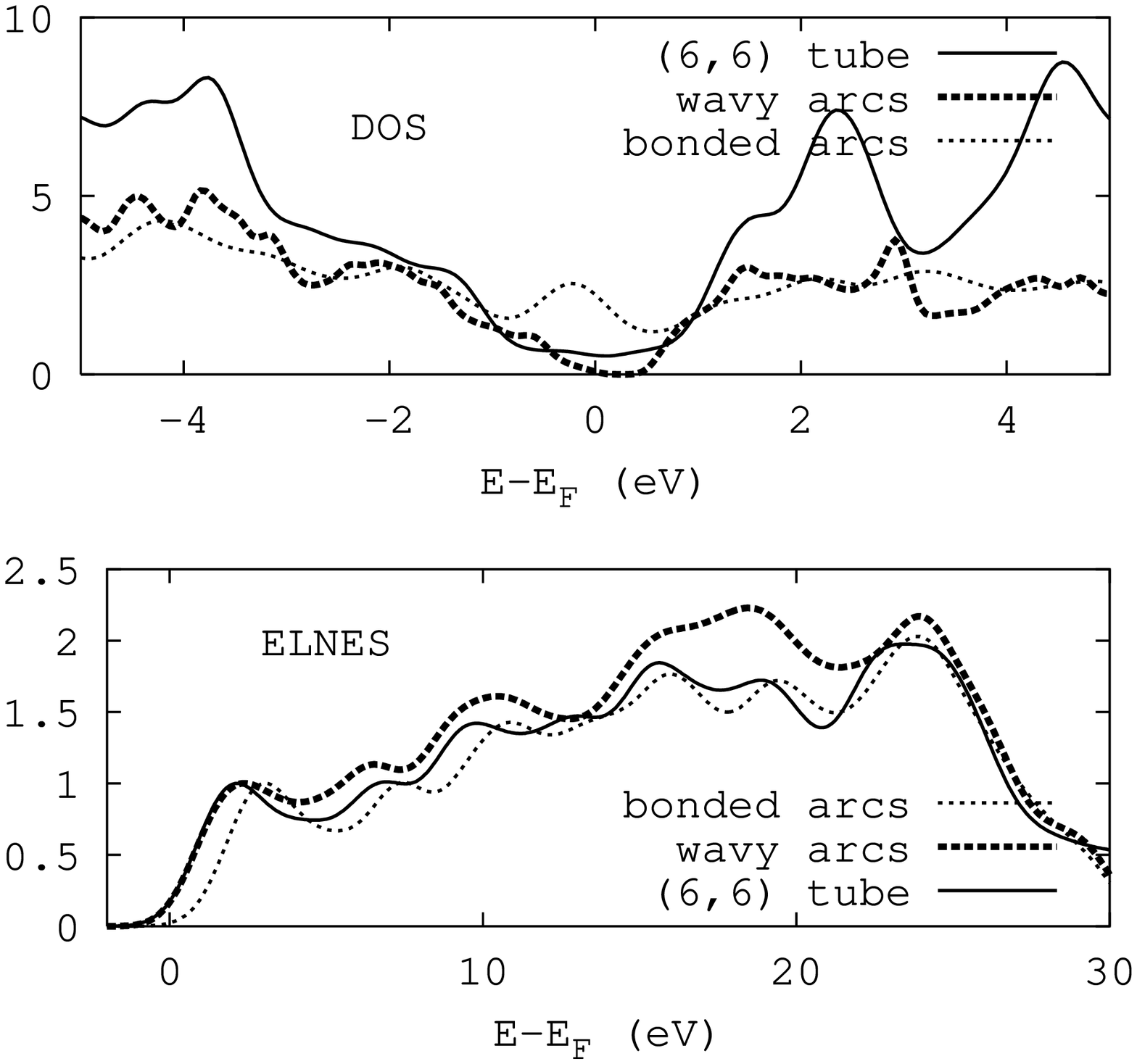}
\end{center}
  \caption{The total DOS and ELNES of the (6,6) tube and its two model arcs.} 
  \label{fig:dos-eels-6,6}
\end{figure}
 \begin{figure}[H]
\begin{center}
\includegraphics[width=3.3in,height=6in,angle=-90]{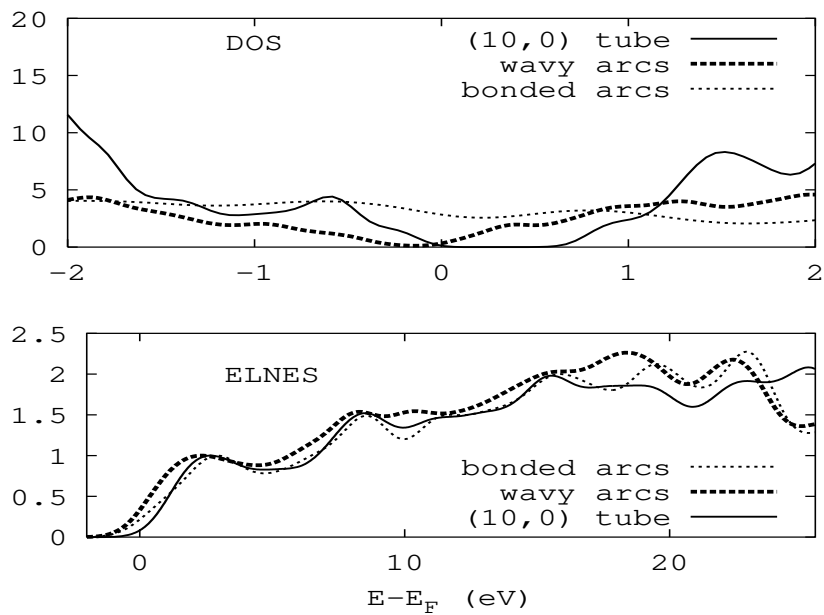}
\end{center}
  \caption{The total DOS and ELNES of the (10,0) tube and its two model arcs.} 
  \label{fig:dos-eels-10,0}
\end{figure}

\section{Conclusions}
We have performed first principles calculations on a series of armchair and zigzag nanotubes of diameters less than 1nm using the all-electron Full-Potential(-Linearised)-Augmented-Plane-Wave method.  We have investigated in detail the effects of curvature, core-hole interaction and orientation on the ELNES of carbon nanotubes. The inclusion of the core-hole effect is very important to reproduce the measured ELNES spectra of carbon nanotubes~\cite{knupfer99,stephan01}. A curvature-induced $\pi-\sigma$ hybridisation is pointed out and shown to have a stronger effect on zigzag tubes than on armchair tubes. Not only does ELNES calculated in the sudden approximation~\cite{suenaga01} fail to 
explain the experimental measurements, an analysis of the $\pi-\sigma$ hybridisation based on it  may be misleading especially for small diameter 
zigzag tubes as they do not show a clear separation between the $\pi^*$ and $\sigma^*$ features. We have shown that the ELNES of carbon nanotubes obeys the dipole selection rule and that the local DOS of the excited atom and the ELNES spectrum may show spikes due to the Van Hove singularity,  which are characteristic of the quasi one-dimensional nature of the nanotubes. The curvature effect is translated into the electron energy loss spectra by the splitting of the $\pi^*$ and the $\sigma^*$ edges. A reduction in the anisotropy, as compared to graphite, is also recorded for these highly curved carbon systems in line with the findings of St\'ephan et al.~\cite{stephan01} although this anisotropy depends on the microscope settings. It is found that the magic convergence and collection semi-angles (1.87, 3.01 mrad) found for graphite gives an electron beam orientation independent ELNES for nanotubes. Therefore, nanotubes can be used as an alternative to graphite as an anisotropic material to determine the magic collection and convergence semi-angles for carbon materials. The advantage of nanotubes over graphite is the ability to explore a large range of focused electron beam orientations. The (l$^\prime$,m$^\prime$) ELNES decomposition has been shown to be very valuable to unambiguously quantify the $\pi-\sigma$ hybridisation in small nanotubes. This decomposition may be a promising technique to study the type of bonding in carbon materials. We have performed calculations on model arcs of tubes and confirmed that ELNES probes the local bonding structures of materials.
\section*{Acknowledgement}
This project was financially supported by the Special Research Fund of the
 University of Antwerp (BOF-NOI)

\end{document}